# Probing Growth-Induced Anisotropic Thermal Transport in CVD Diamond Membranes by Multi-frequency and Multi-spot-size Time-Domain Thermoreflectance

Zhe Cheng[1, a], Thomas Bougher[1, a], Tingyu Bai[3], Steven Y. Wang[3], Chao Li[3], Luke Yates[1], Brian M. Foley[1], Mark Goorsky[3], Baratunde A. Cola[1, 2, b], Samuel Graham[1, 2, b]

[1]. George W. Woodruff School of Mechanical Engineering, Georgia Institute of Technology, Atlanta, Georgia 30332, USA

[2]. School of Materials Science and Engineering, Georgia Institute of Technology, Atlanta, Georgia 30332, USA

[3]. Materials Science and Engineering, University of California, Los Angeles, Los Angeles, CA, 91355, USA

[a.] These authors contributed equally.

[b.] Corresponding Emails: sgraham@gatech.edu. Or cola@gatech.edu.

# Abstract

The maximum output power of GaN-based high-electron mobility transistors is limited by high channel temperature induced by localized self-heating which degrades device performance and reliability. With generated heat fluxes within these devices reaching magnitude close to ten times of that at the sun surface, chemical vapor deposition (CVD) diamond is an attractive candidate to aid in the extraction of this heat in order to keep the operating temperatures of these high power electronics as low as possible. Due to the observed inhomogeneous structure, CVD diamond membranes exhibit a 3D anisotropic thermal conductivity which may result in significantly different cooling performance from expected in a given application. In this work, time domain thermoreflectance (TDTR) is used to measure the thermal properties of an 11.8-µm CVD diamond membrane from its nucleation side. Starting with a spot size diameter larger than the thickness of the membrane, measurements are made at various modulation frequencies from 1.2 MHz to 11.6 MHz to tune the heat penetration depth, and subsequently the part of diamond sampled by TDTR. We divide the membrane into ten sublayers and assume isotropic thermal conductivity in each sublayer. From this, we observe a 2D gradient of the depth-dependent thermal conductivity for this membrane. By measuring the same region with a smaller spot size at multiple frequencies, the in-plane and cross-plane thermal conductivity are extracted respectively. Through this use of multiple spot sizes and modulation frequencies, the 3D anisotropic thermal conductivity of CVD diamond membrane is experimentally obtained by fitting the experimental data to a thermal model. This work provides insight toward an improved understanding of heat conduction inhomogeneity in CVD polycrystalline diamond membrane that is important for applications of thermal management of high power electronics.

## 1. Introduction

Data transmission in modern communication and radar systems which utilize microwave devices requires a certain amount of energy per bit. With data rates from 40 to 100 giga-bits per second (Gbit/s), the ubiquitous deployment of the 4$^{th}$ generation (4G) communication networks and 5$^{th}$ generation (5G) networks around the corner has driven research into high-power GaN-based high-electron mobility transistors (HEMTs).[1] However, localized self-heating within the HEMTs has proven to be an issue at the power densities demanded for various applications, causing high channel temperatures that degrade device performance and reliability.[2] The heat flux can be more than ten times as large as that at the sun surface.[3] Chemical vapor deposition (CVD) diamond with its super-high thermal conductivity is one of the few materials that can provide the thermal relief that is required to keep these devices cool.[4, 5] When growing CVD diamond membranes, diamond nanocrystals are spread on the substrates as seeds. Then diamond continues to grow up from these nucleation sites. With the enlargement of the diamond crystals, they contact mutually and then grow vertically to form a conical structure which makes heat transport in the cross-plane direction different from the in-plane direction. This difference also changes along the cross-plane direction because both in-plane and cross-plane crystal sizes increase with the increasing distance from the nucleation sites. The crystal size in the growth side is much larger than that in the nucleation side. Consequently, polycrystalline diamond membranes show three-dimensional (3D) anisotropy of thermal conduction because of their inhomogeneous structure.[6-10]

The anisotropic heat conduction in CVD diamond has attracted great attention because it affects heat extraction directly. [6, 10-18] However, almost all experimental measurements have been focused on two-dimensional (2D) anisotropy of thermal conduction.[6, 10-18] 3D anisotropy has not been reported before experimentally due to difficulties in thermal measurements. TDTR is a popular noncontact optical pump and probe thermal characterization method used to measure thermal properties of both bulk and nanostructured meatrials.[19-24] A modulated pump beam heats a to-be-measured sample periodically and a delayed beam probes temperature decay of the sample surface by measuring its thermoreflectance. By fitting the experimental signal picked up by a lock-in amplifier to a multi-layer thermal model, thermal properties of the sample can be extracted. In TDTR measurements, the distance heat penetrates into the surface depends on the modulation frequency and the thermal diffusivity of the sample. By tuning the modulation frequency, we can infer the thermal properties of the sample with different penetration depths, which provides an excellent nondestructive way to explore the 3D anisotropy of inhomogeneous CVD diamond membranes.

In this work, we measured thermal properties of a CVD diamond membrane from the nucleation side by multi-frequency and multi-spot-size TDTR. Following deposition of the diamond film, the supporting substrate was etched selectively to fabricate suspended diamond membranes.[9] The membranes were coated with Al layers as transducers for TDTR measurements. Measurements with modulation frequencies from 1.2 to 11.6 MHz were performed with a pump radius of 19.8 µm to measure the thermal properties in the cross-plane direction. On the same spot, measurement with modulation frequencies from 3.6 MHz to 6.3 MHz were performed with a pump radius of 5.28 µm to extract information for the in-plane and cross-plane thermal conductivity. After measuring the grain size distribution of nucleation and growth sides with transmission electron microscope (TEM), a thermal conductivity model was used to understand the experimental results and obtain the 3D anisotropic thermal conductivity. This work provides insight towards an improved understanding of heat conduction inhomogeneity in CVD polycrystalline diamond membranes for applications of heat dissipation in high power electronics.

2. Sample description and characterization

In this work, the diamond membrane was grown on a silicon wafer via CVD by Element Six Company with diamond thicknesses of 11.8 µm. The definition of the growth and nucleation sides are shown in Figure 1 (a). The nucleation side is adjacent to the silicon substrate and the growth side is in contact with the vapor during the growth process. The silicon substrate was etched away with plasma and both the growth and nucleation sides of the diamond were coated with an Al thin film to serve as the transducer layer for subsequent TDTR measurements. The Al thicknesses were determined by a picosecond acoustic method in the TDTR experiments as 103 nm and 218 nm on the growth and nucleation sides, respectively.

The Al layer on the growth side was deposited by E-beam evaporation at the Naval Research Lab. Its thermal conductivity is estimated at 175 W/m-K by measuring the electrical conductivity and applying the Wiedemann-Franz law. For metallic nanostructures, it still remains controversy whether Wiedemann-Franz law holds exactly.[25-28] Therefore, we deposited thick Al layer (218 nm) on the nucleation side by sputtering deposition and obtained high sensitivity for Al thermal conductivity. The Al thermal conductivity is fitted as 152 W/m-K by multi-frequency TDTR measurements and fixed in the subsequent anisotropy thermal model. We calculated the steady-state temperature rises in these measurements and found all of them are very small (less than 1 K). The sensitivity of Al-diamond interface thermal conductance in the growth side is very small so we fix it as 100 MW/m$^2$-K in our data fittings.[29, 30] The pump and probe beam radii measured by a DataRay scanning slit beam profiler are 19.8 μm and 7.3 μm when using the 5X objective; 5.28 μm and 2.1 μm for the 20X objective.

The x-ray diffraction 2θ:ω scan was performed on a Bruker JV D1 diffractometer with Cu Kα$_1$ radiation and a parallel beam source. The acceptance angle for the diffracted beam was ~ 0.36° The sample was mounted vertically on a background-free stage. In the 2θ:ω scan, ω was offset by 10º from the surface orientation of Si substrate in order to avoid Si. Figure 1(b) shows the XRD pattern of the diamond membrane coated with a layer of Al. Peaks of both Al and diamond showed up in the pattern. Additionally, Focused Ion Beam (Nova 600 FIB) was used to prepare plan view samples before they were characterized by a Titan 300 S/TEM (FEI) to obtain the grain morphology of the nucleation and growth sides of the diamond membrane. The scanning transmission electron microscopy (STEM) mode in TEM with High Angle Annular Dark Field (HAADF) detector reveals contrast from different grains. Figure 2 shows the plan view TEM images of the grain structure on the nucleation and growth sides. The dimensions of yellow squares in Figure 2(a) and (b) are 1.57 μm×1.57 μm and 6.27 μm×6.27 μm, respectively. The grain size has been measured along the four lines within the square area and the average value has been calculated for each side. The grain size of the nucleation and growth sides of the diamond membrane were quantified [ASTM E-112. Standard test methods for determining average grain size. (2010)]. A summary of the grain size distributions is shown in Figure 3. The grain size of the growth side is much larger than that of the nucleation side. For the nucleation side, most of the grain sizes are in the range of 60-120 nm with an average grain size of 92 nm.

For the growth side, most of the grain sizes are in the range of 500-1500 nm with an average value of 1360 nm.

## 3. Results and discussion

### 3.1 Gradient thermal conductivity model

To understand the inhomogeneous structure and thermal properties of the CVD diamond membrane, a gradient thermal model similar to that presented by Sood, *et al.* is applied.[6] A schematic diagram for the nature of crystal growth in CVD diamond membranes is shown in Figure 4. Starting at the seed layer, grains are assumed to grow laterally in size as the thickness of the film increases. As they grow, adjacent grains will contact and compete to continue their growth process. The thickness of the diamond membrane is 11.8 μm, which for the model is divided into ten sublayers with different isotropic thermal conductivity ($\kappa_1$-$\kappa_{10}$). In CVD diamond membranes, in-plane grain size increases approximately linearly with the distance from the nucleation interface when diamond thickness is less than 100 μm.[6, 31] In-plane grain size is $L_{in}(z) = d_0 + \alpha * z$ where $d_0$ is the in-plane crystal size in the nucleation side, $\alpha$ is a constant, and $z$ is the distance from the nucleation interface. Cross-plane grain sizes are much larger than in-plane grain sizes. Here, for simplicity, we assume the diamond is isotropic in each single sublayer and, to the first order approximation, define an effective grain size as $L_{eff}(z) = A * L_{in}(z)$, here the constant $A$ includes the effect of anisotropy, phonon transmission, and specularity. We neglect defect scatterings and only consider grain boundary scatterings and phonon-phonon scatterings as the dominant scattering sources which limit the phonon mean free path. So, the phonon mean free path after considering size effect according to Matthiessen's rule is

$$\lambda(z) = \left(\frac{1}{L_{eff}(z)} + \frac{1}{\lambda_{bulk}}\right)^{-1}, \tag{1}$$

Here, $\lambda_{bulk}$ is the phonon mean free path in bulk diamond. In diamond, 80% of heat is carried by phonons with mean free path from 550 nm to 3400 nm.[32] Similar to Ref.6, we take 1 μm as the frequency independent bulk phonon mean free path (gray approximation) where the phonon free path accumulation function shows the steepest increase. Then the thermal conductivity after considering size effect is

$$\kappa_{size}(z) = \frac{\lambda(z) * \kappa_{bulk}}{\lambda_{bulk}}, \tag{2}$$

Here, $\kappa_{bulk}$ is the thermal conductivity of single crystal diamond. We take the theoretical value of 3000 W/m-K based on first-principle calculation.[6, 32] The thermal interface resistance between grains also needs to be considered. The corresponding thermal conductivity is

$$\kappa(z) = \left(\frac{1}{\kappa_{size}(z)} + \frac{R_{gb}}{L_{eff}(z)}\right)^{-1}, \tag{3}$$

Here, thermal interface resistance $R_{gb}$ takes the value of 0.1 m²K/GW according to simulations in Ref.[33, 34] and our data fitting which will discussed later.

Multi-frequency measurements on the same spot on the nucleation side were conducted from 1.2 MHz to 11.6 MHz with 5X objective. If isotropy is assumed when fitting the data, we can obtain frequency dependent effective thermal conductivity as shown in Figure 5. The effective thermal conductivity decreases with increasing frequency. At low modulation frequencies, the penetration depth of the thermal wave is quite large and results in the sampling of more high thermal conductivity diamond further from the nucleation side of the membrane, thereby yielding a larger effective thermal conductivity.

For the gradient thermal model discussed previously, constants $A$ and $R_{gb}$ are unknown parameters which can be determined by fitting the model to the frequency-dependent effective thermal conductivity. An iterative fitting procedure is employed where an initial guess is made for the two values of $A$ and $R_{gb}$. Using this initial guess for $A$ and $R_{gb}$, the ten-layer gradient thermal conductivity is calculated and used to generate a theoretical TDTR $-V_{in}/V_{out}$ curve. We then fit this theoretical curve with an isotropic 3-layer (Al-Diamond-Al) TDTR model to obtain an effective thermal conductivity for the sampled volume of the membrane. After comparing this calculated value to the measured effective thermal conductivity, we adjust the values of $A$ and $R_{gb}$. These steps are repeated until the calculated and measured effective thermal conductivities converge, resulting in fit values of $A= 2$ and $R_{gb} = 0.1$ m$^2$ K/GW. The fitted value of $A= 2$ indicates the effective mean free path is twice of the in-plane crystal size. This is reasonable because the cross-plane crystal size is much larger than the in-plane crystal size. $R_{gb}$ is also close to the simulation values reported in Ref. [33, 34]. The good agreement between the measured effective thermal conductivities and the calculated gradient-model values using $A= 2$ and $R_{gb}= 0.1$ m$^2$ K/GW are depicted in Figure 5.

It should be noted that the sensitivity of the Al-diamond interface conductance at the nucleation side when fitting the experimental data is very large at all thermal modulation frequencies (comparable to that of the diamond thermal conductivity for 1.2 MHz, about five times as large as that of the diamond thermal conductivity for 11.6 MHz). When fitting the TDTR data at each modulation frequency for both the Al-diamond interface conductance and the effective thermal conductivity of the diamond, the extracted value for the Al-diamond interface conductance consistently fit at 91 ±2 MW/m$^2$-K with the small uncertainty attributed to the fact that the measurements at each modulation frequency were made at the same spot on the film. This is an important point in that even with the larger sensitivity to this parameter compared to the diamond thermal conductivity, the value for the Al-diamond interface conductance is independent of the thermal modulation frequency, thereby reinforcing the observed relation between the modulation frequency and the effective diamond thermal conductivity.

Figure 6 shows the local thermal conductivity of the ten layers in the gradient thermal conductivity model assuming the fitted values for $A$ and $R_{gb}$. The local thermal conductivity increases with the distance from the nucleation interface. Near the nucleation interface, grain sizes are small and phonon scattering at the grain boundaries is the dominant mechanism that limits the phonon mean free path. The local thermal conductivity as a function of distance from the nucleation side of the film exhibits a non-linear trend. Near the surface of the growth side, grain sizes are large and phonon-phonon (Umklapp) scattering is the dominant mechanism which limits the phonon mean free path. However, as we move closer to the nucleation side of the membrane, the local thermal conductivity decreases as the smaller and smaller grains cause the phonons to scatter more often with grain boundaries than with themselves. All the above analysis is based on a first order approximation and a more detailed analysis of the 3D anisotropic thermal conductivity will be discussed later.

**3.2 Anisotropic thermal conductivity measured by different spot sizes**

When the TDTR beam spot size is much larger than the penetration depth, heat transfers one dimensionally along the cross-plane direction and the sensitivity of the TDTR data for fitting the cross-plane thermal conductivity is much larger than that of in-plane thermal conductivity, as shown in Figure 7. When using the large spot sizes provided with the 5x objective (19.8 μm and 7.3 μm radii for the pump and probe, respectively), the sensitivity of $\kappa_z$ is much larger than that of $\kappa_r$, allowing us to extract cross-plane thermal conductivity easily from the 5X objective measurement. At the same spot on the same sample with the same modulation frequency, we perform another TDTR measurement with a 20X objective (5.28 μm and 2.1 μm radii for the pump and probe, respectively). When using the smaller spot sizes produced by the 20x, heat transfers more radially and the sensitivity of the TDTR measurement to the in-plane thermal conductivity increases while that to cross-plane thermal conductivity decreases, as shown in Figure 7. Using these two different sets of spot sizes to take separate TDTR scans at the same location on the sample, we can iteratively determine both the in-plane and cross-plane thermal conductivities by finding values that fit both independent TDTR measurements. Our procedure is as follows; we first fit the cross-plane thermal conductivity using data collected with the 5X objective. We then insert this value for cross-plane thermal conductivity into the data fitting as a fixed parameter and then fit for the in-plane thermal conductivity using TDTR data taken at the same location with the 20X objective. This value for the in-plane thermal conductivity is then input to and held fixed in the data fitting and the 5X TDTR data is then fitted for the cross-plane conductivity, and the iterative procedure is repeated until both the in-plane and cross-plane thermal conductivities fit well in both data sets. Therefore, with different spot sizes measured at the same spot with the same modulation frequency, we obtain both in-plane and cross-plane thermal conductivity.

In this work, we performed TDTR measurements with 5X and 20X objectives for frequencies of 3.6 MHz and 6.3 MHz. For smaller frequencies, the sensitivity of the TDTR data to the in-plane and cross-plane thermal conductivities with the 5X objective are comparable and neither the cross-plane nor in-plane thermal conductivities can be extracted independently. For larger frequencies, the sensitivity of the data to the in-plane thermal conductivity when using the 20X objective is too small to obtain accurate in-plane thermal conductivity.[35] The measured cross-plane and in-plane thermal conductivity are shown in Table 1. The cross-plane thermal conductivity of 3.6 MHz and 6.3 MHz are 1296 and 1182 W/m-K, while the corresponding in-plane thermal conductivity are 620 and 531 W/m-K, respectively. While the values for the in-plane thermal conductivity compare quite well between the two modulation frequencies used (<10 % difference), the cross-plane thermal conductivities differ by as much as 17%. This is consistent with the explanation provided earlier in the manuscript where the smaller penetration depth associated with the higher modulation frequency results in a greater fraction of the volume near the nucleation side being sampled, yielding lower effective cross-plane thermal conductivity. It should again be mentioned that the fitted values for the Al-diamond interface conductance are very consistent with each other.

### 3.3 3D anisotropic thermal conductivity

For 3D anisotropic thermal conduction in the CVD diamond membrane, we used the thermal model in Ref. 6. In-plane grain size ($d_r$) is modeled as $d_r(z) = d_0 + \alpha * z$, where $d_0$ is the crystal size at the nucleation interface, $\alpha$ is a constant, and $z$ is the distance from the nucleation interface. Cross-plane crystal size ($d_z$) is given by

$$\frac{d_z(z)}{L} = (\frac{z}{L} + \frac{\beta}{\alpha}) * \left[ \frac{1}{\log(g)} \log(\frac{\alpha+\beta}{\alpha z/L + \beta}) + 1 \right] - \frac{\beta}{\alpha} \tag{4}$$

where $\beta = d_0/L$ and $g$ is the inverse survival rate, which in this work we fix as 2.[6] Taking into consideration several types of phonon scattering mechanisms, including phonon-phonon, as well as in-plane and cross-plane grain boundary scattering, the effective mean free path in the cross-plane and in-plane directions can be obtained by[6, 36, 37]

$$\lambda_{z,r}^{-1}(z) = \frac{1}{\lambda_{bulk}} + \frac{1}{B * d_{z,r}(z)} + \frac{1}{C * d_{r,z}(z)} \tag{5}$$

where $B$ and $C$ are constants that are related to the probability of phonon transmission $t$ and the specularity parameter $p$. Assuming a gray (frequency-independent) approximation with regards to these parameters, $B = 0.75t/(1-t)$ and $C = (1+p)/(1-p)$ represent the cases where heat transport is limited by scattering that is perpendicular to interfaces ($t$ effect) and parallel to interfaces ($p$ effect). After obtaining the in-plane and cross-plane effective phonon mean free paths, the in-plane and cross-plane thermal conductivity can be obtained by inserting them into Equations (2) & (3), assuming $R_{gb}$ = 0.1 m²K/GW. Finally, we can fit $t$ and $p$ using TDTR measurements collected at 3.6 MHz and 6.3 MHz, using both the 5X and 20X objectives. The diamond membrane is divided into ten layers, as shown in Figure 4. For each layer, we can obtain the in-plane and cross-plane thermal conductivity through an iterative method similar to that described previously; first we a guess for the values of $t$ and $p$, then we generate a theoretical TDTR $-V_{in}/V_{out}$ curve which we can compare to the experimental data and then revise our guesses for $t$ and $p$ to repeat the process. Finally, we get $t$ as 0.56 and $p$ as 0.33 which are reasonable for relatively high quality diamond grain boundaries. Figure 8 shows the good agreement between TDTR experimental data and theoretical curves. The 3D anisotropic thermal conductivity (change in the in-plane and cross-plane thermal conductivity with the distance from the nucleation interface) are shown in Figure 9. Both the in-plane and cross-plane thermal conductivity increase significantly with increasing distance from the nucleation interface because thermal transport goes from being dominated by grain boundary scattering to phonon-phonon scattering as the grains get larger.

In this work, we attribute the frequency and spot-size dependent thermal conductivity to anisotropic structure of CVD diamond. We noticed that, in TDTR measurements, ballistic thermal transport results in reduced thermal conductivity when laser spot size or penetration depth are smaller or comparable to phonon mean free path because phonons with mean free path larger than spot size or penetration depth does not contribute to thermal conduction.[38-40] In bulk diamond, at room temperature phonons with mean free path from 550 nm to 3400 nm contribute to 80% of heat conduction.[32] In CVD diamond, especially the nucleation side of the membrane, the small crystal size limits the phonon mean free path, making ballistic thermal transport more difficult to be observed. In our measurements, the root-mean spot-sizes for 5X and 20X are 14.9 µm and 4.0 µm, respectively. In Ref. 36, reduction in thermal conductivity can be observed only if the root-mean spot-size is smaller than 2 µm for bulk diamond. This limit would be much smaller for CVD polycrystalline diamond membrane. So, ballistic thermal transport should not show up in measurements with these spot sizes. In terms of modulation frequency, we can estimate the penetration depth with the formula $\sqrt{\alpha_t/\pi f}$, here $\alpha_t$ is thermal diffusivity of diamond and *f* is modulation frequency in TDTR measurements. The smallest penetration depth in this paper is around 3.5 µm, which is much larger than possible phonon mean free path in the CVD diamond. Therefore, ballistic thermal transport does not affect our TDTR measurements to probe the inhomogeneous thermal conduction in CVD diamond.

## 4. Conclusion

In order to cool high power electronics effectively and mitigate premature degradation or failure, high thermal conductivity materials that can be integrated at the die level are absolutely critical. CVD diamond is one of the few materials that has provided a disruptive advancement in the field of near-junction thermal relief solutions. However, because of the inhomogeneous structure arising during the CVD growth process, the thermal conductivities of these diamond films are highly anisotropic, affecting heat dissipation significantly. In this work, TDTR was used to measure thermal properties of an 11.8-μm CVD diamond membrane from its nucleation side. By changing modulation frequencies from 1.2 MHz to 11.6 MHz to tune the heat penetration depth and subsequently the part of diamond sampled by TDTR, we obtained a 2D gradient thermal conductivity and observed frequency/depth dependent effective thermal conductivity. In addition, by measuring the same spot on the sample with different spot sizes, we were able to independently determine the in-plane and cross-plane thermal conductivity. Through a combination of using multiple spot sizes and multiple modulation frequencies, (3.6 MHz and 6.3 MHz), the full 3D anisotropic thermal conductivity of a CVD diamond membrane was experimentally obtained. This work provides insights in measuring and understanding heat conduction inhomogeneity in CVD polycrystalline diamond membrane for applications of thermal management of high power electronics.

References

1. Y.-F. Wu, M. Moore, A. Saxler, T. Wisleder and P. Parikh, 2006.

2. J. Cho, Z. Li, E. Bozorg-Grayeli, T. Kodama, D. Francis, F. Ejeckam, F. Faili, M. Asheghi and K. E. Goodson, *IEEE Transactions on Components, Packaging and Manufacturing Technology*, 2013, **3**, 79-85.

3. F. Faili, N. L. Palmer, S. Oh and D. J. Twitchen, presented in part at the ITherm Orlando, Florida, US, 05/2017, 2017.

4. J. W. Pomeroy, M. Bernardoni, D. Dumka, D. Fanning and M. Kuball, *Applied Physics Letters*, 2014, **104**, 083513.

5. L. Yates, A. Sood, Z. Cheng, T. Bougher, K. Malcolm, J. Cho, M. Asheghi, K. Goodson, M. Goorsky and F. Faili, 2016.

6. A. Sood, J. Cho, K. D. Hobart, T. I. Feygelson, B. B. Pate, M. Asheghi, D. G. Cahill and K. E. Goodson, *Journal of Applied Physics*, 2016, **119**, 175103.

7. J. Anaya, S. Rossi, M. Alomari, E. Kohn, L. Tóth, B. Pécz, K. D. Hobart, T. J. Anderson, T. I. Feygelson and B. B. Pate, *Acta Materialia*, 2016, **103**, 141-152.

8. K. E. Goodson, *Journal of Heat Transfer*, 1996, **118**, 279-286.

9. E. Bozorg-Grayeli, A. Sood, M. Asheghi, V. Gambin, R. Sandhu, T. I. Feygelson, B. B. Pate, K. Hobart and K. E. Goodson, *Applied Physics Letters*, 2013, **102**, 111907.

10. J. Graebner, S. Jin, G. Kammlott, B. Bacon, L. Seibles and W. Banholzer, *Journal of applied physics*, 1992, **71**, 5353-5356.

11. J. Graebner, S. Jin, G. Kammlott, J. Herb and C. Gardinier, *Nature*, 1992, **359**, 401-403.

12. J. Graebner, S. Jin, G. Kammlott, J. Herb and C. Gardinier, *Applied physics letters*, 1992, **60**, 1576-1578.

13. J. Graebner, M. Reiss, L. Seibles, T. Hartnett, R. Miller and C. Robinson, *Physical Review B*, 1994, **50**, 3702.

14. E. V. e. Ivakin, A. V. e. Sukhodolov, V. G. e. Ralchenko, A. V. Vlasov and A. V. Khomich, *Quantum Electronics*, 2002, **32**, 367.

15. A. Sukhadolau, E. Ivakin, V. Ralchenko, A. Khomich, A. Vlasov and A. Popovich, *Diamond and related materials*, 2005, **14**, 589-593.

16. H. Verhoeven, A. Flöter, H. Reiß, R. Zachai, D. Wittorf and W. Jäger, *Applied physics letters*, 1997, **71**, 1329-1331.

17. S. Rossi, M. Alomari, Y. Zhang, S. Bychikhin, D. Pogany, J. Weaver and E. Kohn, *Diamond and Related Materials*, 2013, **40**, 69-74.

18. J. Anaya, S. Rossi, M. Alomari, E. Kohn, L. Tóth, B. Pécz and M. Kuball, *Applied Physics Letters*, 2015, **106**, 223101.

19. T. L. Bougher, L. Yates, C.-F. Lo, W. Johnson, S. Graham and B. A. Cola, *Nanoscale and Microscale Thermophysical Engineering*, 2016, 1-11.

20. T. L. Bougher, J. H. Taphouse and B. A. Cola, 2015.

21. L. Zeng, *Massachusetts Institute of Technology*, 2016.

22. L. Zeng, K. C. Collins, Y. Hu, M. N. Luckyanova, A. A. Maznev, S. Huberman, V. Chiloyan, J. Zhou, X. Huang and K. A. Nelson, *Scientific reports*, 2015, **5**.

23. D. G. Cahill, *Review of scientific instruments*, 2004, **75**, 5119-5122.

24. A. J. Schmidt, X. Chen and G. Chen, *Review of Scientific Instruments*, 2008, **79**, 114902.

25. S. Yoneoka, J. Lee, M. Liger, G. Yama, T. Kodama, M. Gunji, J. Provine, R. T. Howe, K. E. Goodson and T. W. Kenny, *Nano letters*, 2012, **12**, 683-686.

26. Z. Cheng, 2015.

27. Z. Cheng, L. Liu, S. Xu, M. Lu and X. Wang, *Scientific Reports*, 2015, **5**, 10718.
28. Z. Cheng, Z. Xu, S. Xu and X. Wang, *Journal of Applied Physics*, 2015, **117**, 024307.
29. G. T. Hohensee, R. Wilson and D. G. Cahill, *Nature communications*, 2015, **6**, 6578.
30. H.-K. Lyeo and D. G. Cahill, *Physical Review B*, 2006, **73**, 144301.
31. R. B. Simon, J. Anaya, F. Faili, R. Balmer, G. T. Williams, D. J. Twitchen and M. Kuball, *Applied Physics Express*, 2016, **9**, 061302.
32. W. Li, N. Mingo, L. Lindsay, D. A. Broido, D. A. Stewart and N. A. Katcho, *Physical Review B*, 2012, **85**, 195436.
33. H. Dong, B. Wen and R. Melnik, *Scientific reports*, 2014, **4**, 7037.
34. D. Spiteri, J. Anaya and M. Kuball, *Journal of Applied Physics*, 2016, **119**, 085102.
35. P. Jiang, X. Qian and R. Yang, *arXiv preprint arXiv:1704.02358*, 2017.
36. Z. Wang, J. E. Alaniz, W. Jang, J. E. Garay and C. Dames, *Nano letters*, 2011, **11**, 2206-2213.
37. T. Hori, J. Shiomi and C. Dames, *Applied Physics Letters*, 2015, **106**, 171901.
38. Y. K. Koh and D. G. Cahill, *Physical Review B*, 2007, **76**, 075207.
39. R. Wilson and D. G. Cahill, *Applied Physics Letters*, 2015, **107**, 203112.
40. L. Zeng, K. C. Collins, Y. Hu, M. N. Luckyanova, A. A. Maznev, S. Huberman, V. Chiloyan, J. Zhou, X. Huang and K. A. Nelson, *Scientific reports*, 2015, **5**, 17131.

Table 1. Cross-plane and in-plane thermal conductivity measured with different spot sizes for different modulation frequencies.

|  | $\kappa_z$ (W/m-K) | $\kappa_r$ (W/m-K) |
|---|---|---|
| 3.6 MHz | 1296 | 620 |
| 6.3 MHz | 1182 | 531 |

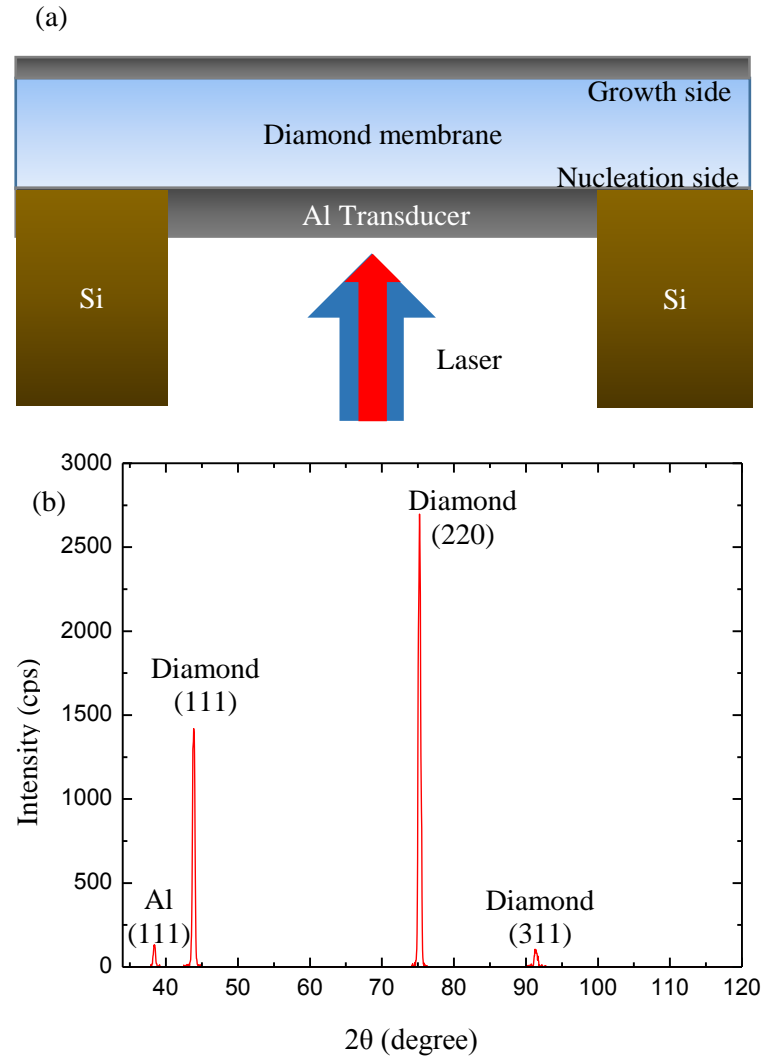

Figure 1. (a) Schematic diagram of the sample structure. The silicon substrate is etched and Al transducer is coated on both the growth and nucleation sides. Pump (blue) beam heats the sample and probe (red) beam measures the temperature variation. (b) XRD pattern for the diamond membrane coated with a layer of Al. Both Al and diamond peaks show up in the pattern.

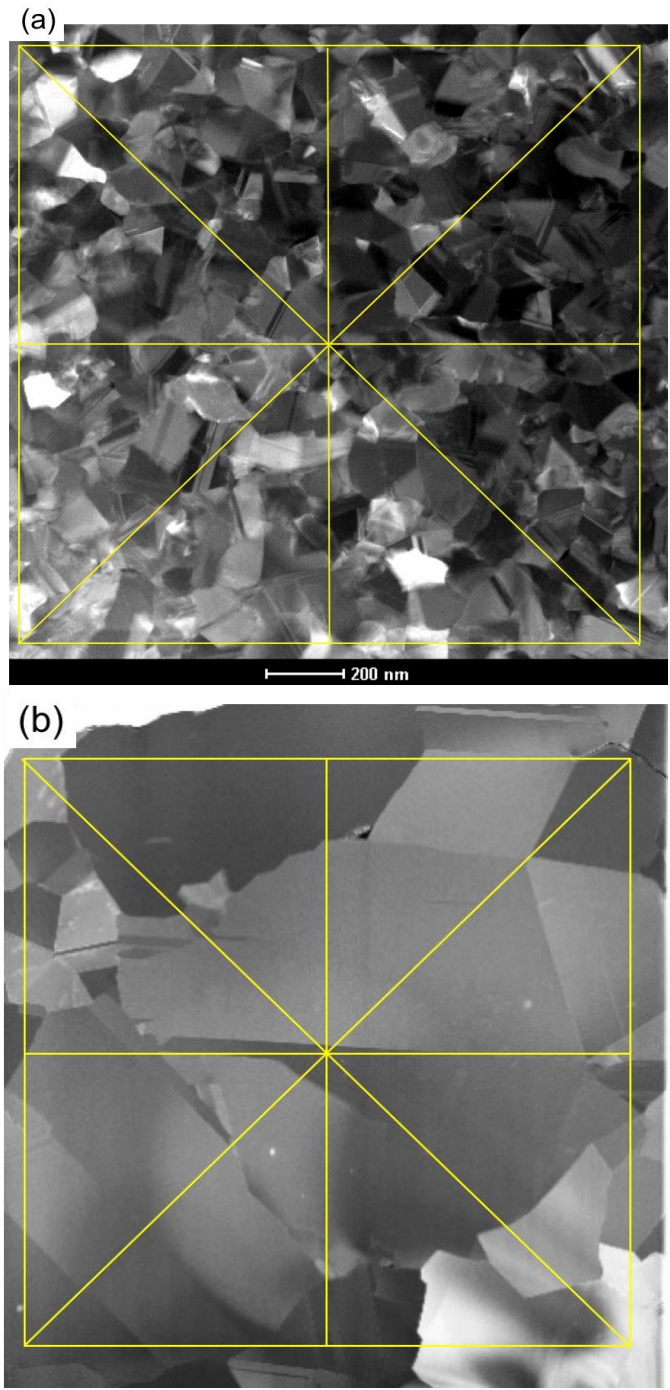

Figure 2. Plan-view grains of nucleation (a) and growth (b) sides of diamond membrane measured by TEM. The dimensions of yellow squares in (a) and (b) are 1.57 μm×1.57 μm and 6.27 μm×6.27 μm, respectively.

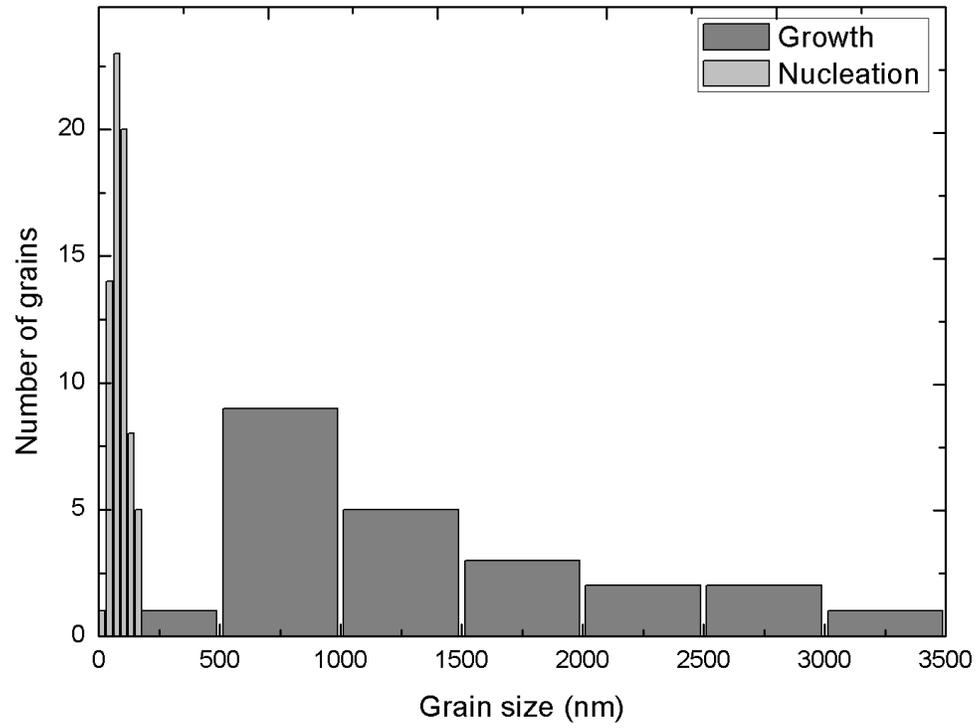

Figure 3. Grain size distributions of nucleation and growth sides of the diamond membrane.

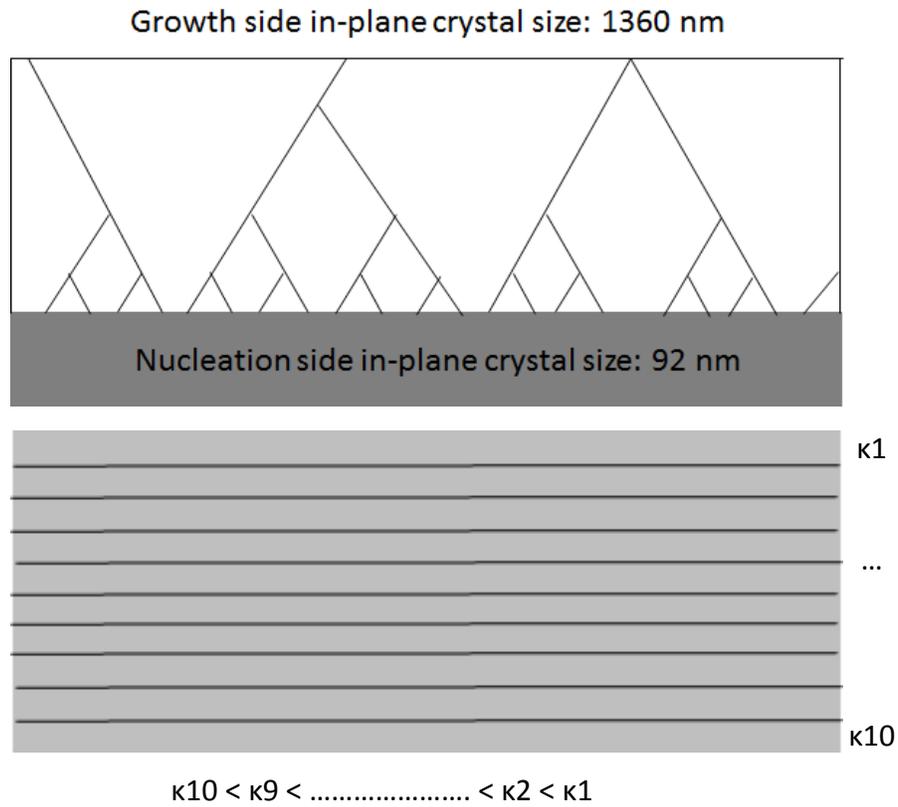

Figure 4. Schematic diagram of crystal growth in CVD diamond. Total thickness of the diamond membrane is 11.8 μm. It is divided into ten sublayers with different isotropic thermal conductivity $κ1, κ2,…, κ10$.

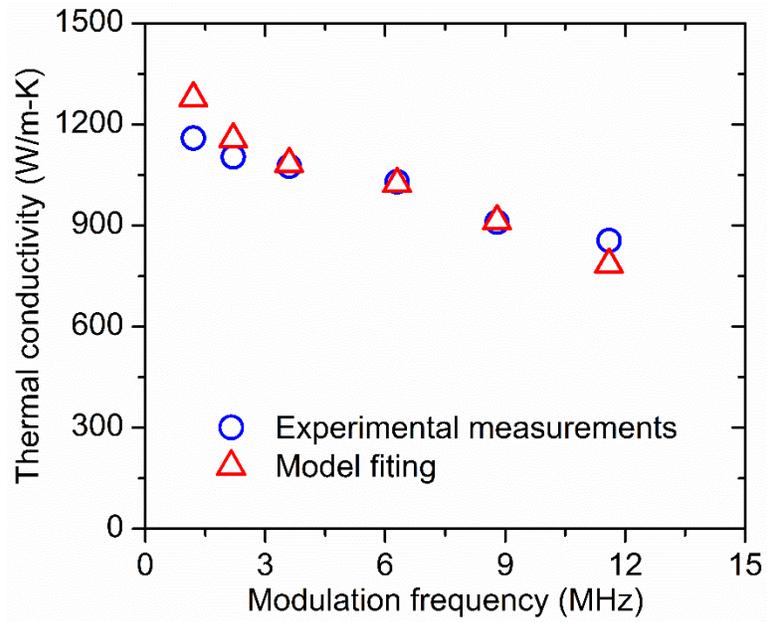

Figure 5. Frequency dependence of thermal conductivity of the diamond membrane.

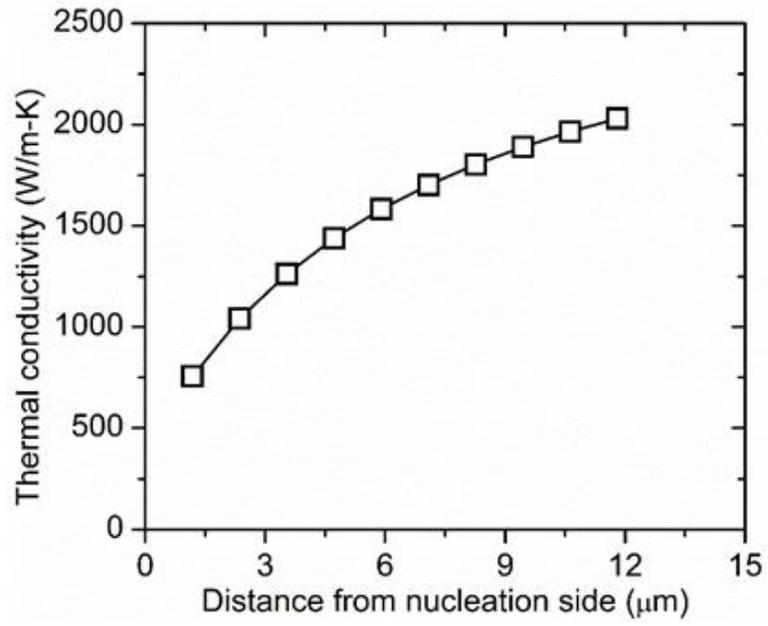

Figure 6. Gradient thermal conductivity of the diamond membrane.

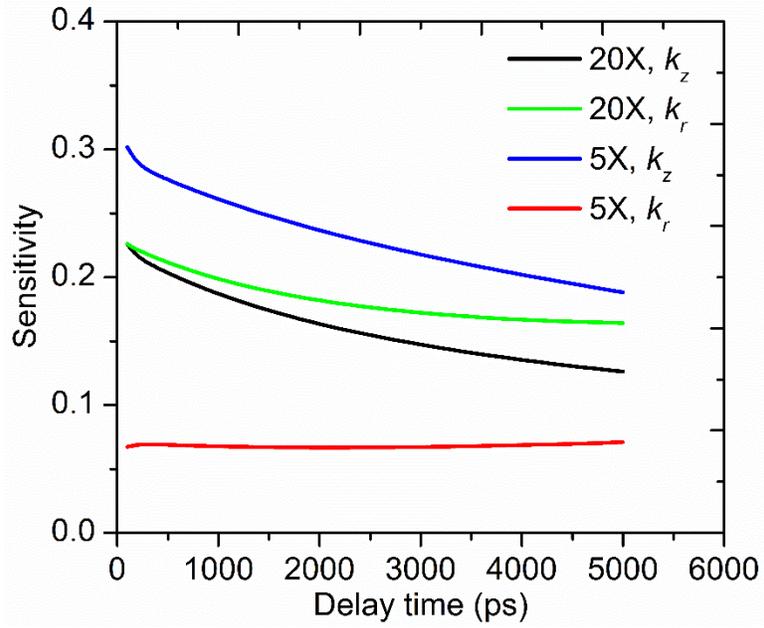

Figure 7. Sensitivity of in-plane and cross-plane thermal conductivity with 20X and 5X objectives. The pump and probe radii are 5.28 μm and 2.1 μm for 20X objective and 19.8 μm and 7.3 μm for 5X objective.

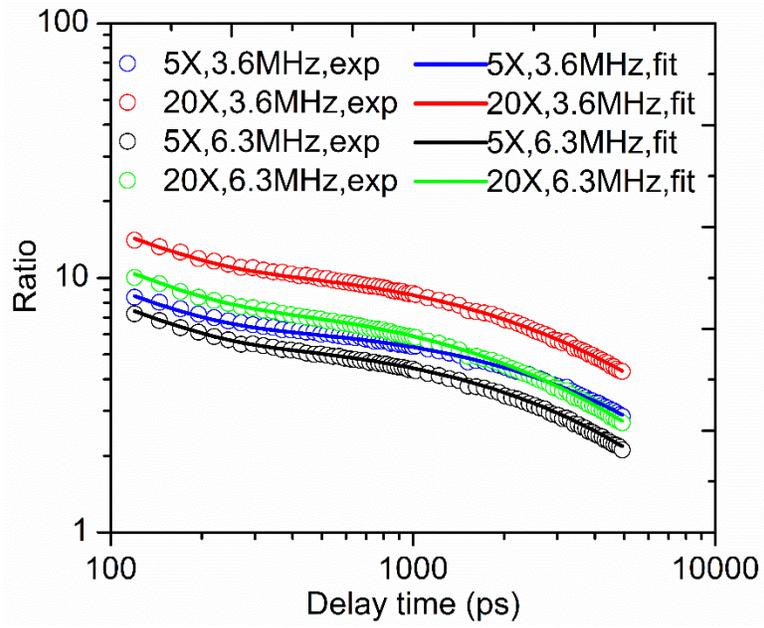

Figure 8. Experimental measurements are fitted with a theoretical model.

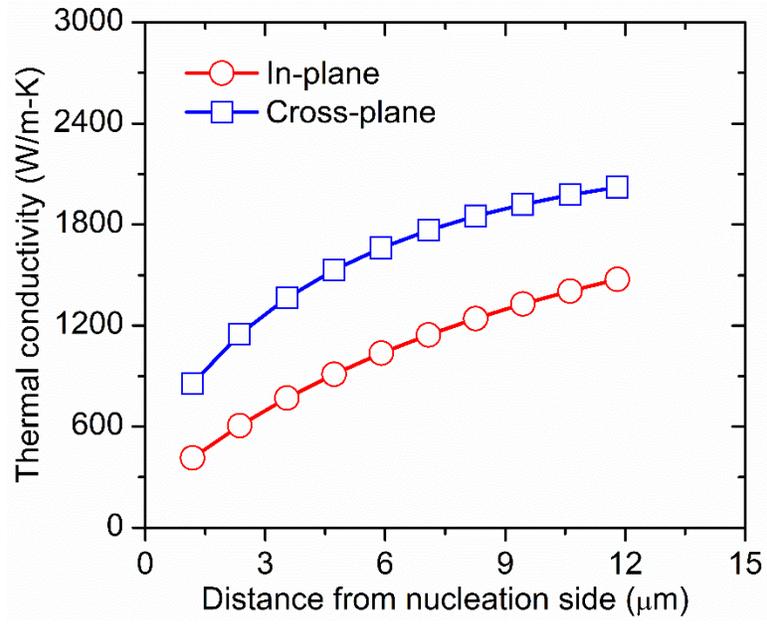

Figure 9. In-plane and cross-plane thermal conductivity along the cross-plane direction.